\documentclass[prb,twocolumn,amsmath,amssymb,superscriptaddress]{revtex4-1}

\usepackage{graphicx}
\usepackage{dcolumn}
\usepackage{hhline}
\usepackage{bm,color}

\usepackage{float}

\usepackage{amsmath}

\usepackage{epstopdf}

\begin{document}


\title{Soft tilt and rotational modes in the hybrid improper ferroelectric Ca$_{3}$Mn$_{2}$O$_{7}$}

\author{A. Glamazda}
\affiliation{Department of Physics, Chung-Ang University, Seoul 156-756, Republic of Korea}

\author{D. Wulferding}
\affiliation{Laboratory for Emerging Nanometrology (LENA), TU Braunschweig, 38106 Braunschweig, Germany}

\author{P. Lemmens}
\affiliation{Laboratory for Emerging Nanometrology (LENA), TU Braunschweig, 38106 Braunschweig, Germany}
\affiliation{Inst. for Condensed Matter Physics, TU Braunschweig, D-38106 Braunschweig, Germany}

\author{B. Gao}
\affiliation{Rutgers Center for Emergent Materials and Department of Physics and Astronomy, Rutgers University, Piscataway, New Jersey 08854, USA}

\author{S.-W. Cheong}
\affiliation{Rutgers Center for Emergent Materials and Department of Physics and Astronomy, Rutgers University, Piscataway, New Jersey 08854, USA}

\author{K.-Y. Choi}
\email[]{kchoi@cau.ac.kr}
\affiliation{Department of Physics, Chung-Ang University, Seoul 156-756, Republic of Korea}


\begin{abstract}
Raman spectroscopy is employed to probe directly the soft rotation and tilting modes, which are two primary order parameters predicted in the hybrid improper ferroelectric material Ca$_3$Mn$_2$O$_7$.  We observe a giant softening of the 107-cm$^{-1}$ octahedron tilting
mode by 26~cm$^{-1}$, on heating through the structural transition from a ferroelectric to  paraelectric orthorhombic phase.
This is contrasted by a small softening of the 150-cm$^{-1}$ rotational mode by 6~cm$^{-1}$.
In the intermediate phase, the competing soft modes with different symmetries
coexist, bringing about many-faceted anomalies in spin excitations and lattice vibrations.
Our work demonstrates that the soft rotation and tilt patterns, relying on a phase-transition path,
are a key factor in determining ferroelectric, magnetic, and lattice properties of Ca$_3$Mn$_2$O$_7$.
\end{abstract}

\maketitle

\section{\label{sec:level1}Introduction}
Hybrid improper ferroelectrics (HIFs) are currently of great interest because they hold promise for the electric field control of nonpolar order parameters as well as for the realization of room-temperature multiferroelectricity~\cite{Bousquet,Benedek11,Rondinelli,Benedek12}. In HIFs  a polarization  is generated by a combination of
rotation ($X^{+}_2$) and tilt ($X^{-}_3$) distortions of the oxygen octahedra (see Fig.~1). In this case, a primary order parameter is given by the amplitude of hybrid $ X^{+}_2 \oplus  X^{-}_3$ distortions, giving rise to rotational (designated R) and tilting (T) soft modes.  Despite their technological importance, little is known about their impact on spin, lattice, and ferroelectric dynamics.

The HIF was observed in perovskite superlattices  and  $n=2$ Ruddlesden-Popper materials Ca$_3$(Ti,Mn)$_2$O$_7$ \cite{Bousquet,Benedek11,Harris,Mulder,Rondinelli,Varignon,Xu,Oh,Birol,Senn15,Huang,Huang16,Lee}. The prototypical HIF Ca$_3$Mn$_2$O$_7$ consists of an alternating arrangement of the CaO rocksalt layers and CaMnO$_3$ bilayers along the  $c$ axis as shown in Fig. 1(a).
The Ca$_3$Mn$_2$O$_7$ compound is known to undergo a first-order phase transition from a
high-$T$ tetragonal to a low-$T$ orthorhombic phase through a certain intermediate phase~\cite{Senn15,Lobanov,Gao}.
An early study has shown that at room temperature the compound comprises a mixture of orthorhombic ($Cmc2_1$) and tetragonal ($I4/mmm$)
phases, which evolves into a single $Cmc2_1$ phase at low temperature~\cite{Lobanov}.
However, this structural model was questioned by later refined structural studies~\cite{Senn15,Gao}.
With decreasing temperature the high-$T$ tetragonal $I4/mmm$ structure undergoes a transition to an intermediate paraelectric orthorhombic ($Acaa$)
and then to a polar  $A2_1am$ symmetry at $T_{\mathrm{S}}=310$~K. As the temperature is further lowered, a G-type antiferromagnetic order sets in  at $T_\mathrm{N}=115$~K~\cite{Lobanov}.

\begin{figure}
\label{figure1}
\centering
\includegraphics[width=6.5cm]{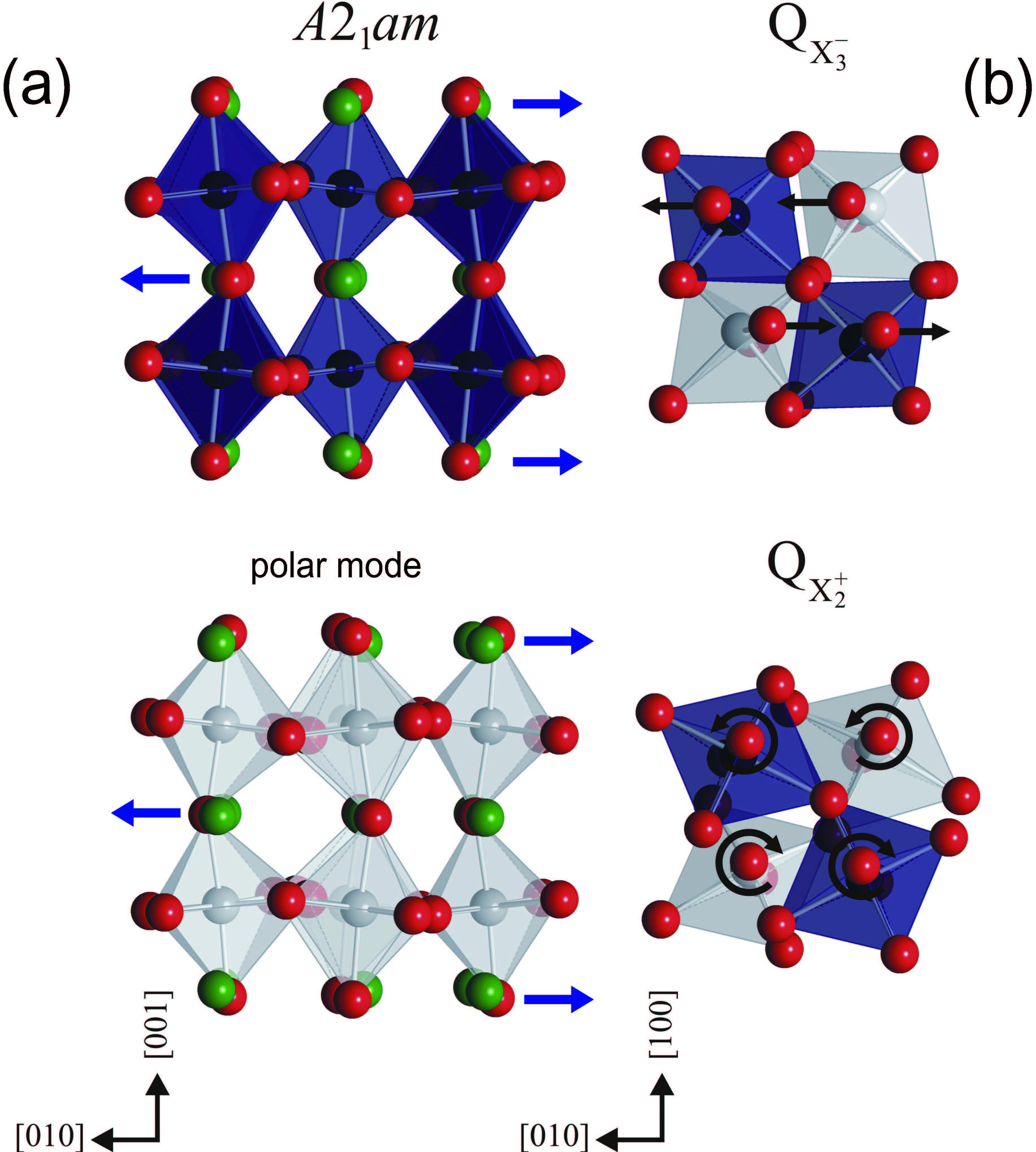}
\caption{(a) Low-temperature crystal structure of Ca$_3$Mn$_2$O$_7$ in its distorted ferroelectric state. Mn ions (gray) are surrounded by oxygen octahedra
(red O ions) with Ca ions (green) interspersed. The arrows depict the polar displacements of Ca ions. (b) Octahedral rotation ($X^{+}_2$) and octahedral tilt ($X^{-}_3$) distortions. }
\end{figure}

Noteworthy is that the intermediate $Acaa$ symmetry is generated by an out-of-phase rotation ($X^{-}_1$) of the MnO$_6$ octahedra  from the undistorted $I4/mmm$ symmetry. In contrast, the ferroelectric $A2_1am$ phase below room temperature is associated with the hybridized distortions of the $X^{+}_2$ rotation in the $xy$ plane and the $X^{-}_3$ octahedral tilt in the $yz$ plane. The transition pathway from  $I4/mmm$ through $Acaa$ to $A2_1am$ symmetries results in
the competition of the $X^{-}_1$ out-of-phase and $X^{+}_2$ in-phase rotations, thereby bringing about peculiar physical
phenomena. Unlike the isostructural Ca$_3$Ti$_2$O$_7$, on the one hand, a polarization is not switchable due to the formation of 90$^{\circ}$ ferroelastic domains stacked alternately along the $c$ axis~\cite{Gao}. On the other hand, uniaxial negative thermal expansion, resulting from  symmetry trapping of the soft T mode, is observed only in the $Acaa$ phase~\cite{Senn15,Senn16}. Thus, tracing the tilt and rotation modes through the phase transitions is indispensable for
the in-depth understanding of the HIF of Ca$_3$Mn$_2$O$_7$.

In this paper, we report a successful observation of the soft octahedral tilt and rotational modes in Ca$_3$Mn$_2$O$_7$ using polarization-resolved Raman spectroscopy. The salient finding is that the antiphase octahedron tilting mode undergoes a giant softening upon heating towards $T_{\mathrm{S}}$, while the rotational mode experiences a small softening.
In the intermediate phase, we further find coexistence of the competing soft modes belonging to distinct phases, which is a main cause of ferroelectric, magnetic, and lattice anomalies.

\section{\label{sec:level2}Experimental Details}
Single crystals of Ca$_3$Mn$_2$O$_7$ were grown using a floating zone method~\cite{Gao}.
For Raman experiments,  samples with dimensions of $5\times5 \times2\, \mbox{mm}^3$ were installed in an evacuated closed-cycle cryostat at $T=5-360$~K. The incoming light was normal to the  $ab$ surface as well as parallel to the $c$ axis for the polarized Raman scattering experiment. The scattered spectra were collected in (quasi)backscattering geometry and were dispersed by a triple spectrometer (Dilor-XY-500) and a micro-Raman spectrometer (Jobin Yvon LabRam) equipped with a liquid-nitrogen-cooled CCD detector. The laser beam ($P=8$~mW) was used to avoid the local heating of the sample.
For the resonance Raman spectra, both solid-state lasers and an Ar-Kr ion laser with the excitation lines of $\lambda= 488 -647$~nm  were employed.

\section{\label{sec:level2}Lattice-dynamical calculations}

\begin{table*}[hth] \caption{\label{tab:table1} List of shell model parameters for the shell-shell and core-shell interactions.}
\begin{ruledtabular}\begin{tabular}{ cccccccc }
 Atom &  X($|e|$) & Y($|e|$) & K(eV/${\AA}$) & Atomic pair & A(eV) & $\rho(\AA)$ & C(eV$\AA^6$) \\ \hline
 Ca      & 2  &  0   & 0  &  Ca-O  & 1090.4  & 0.3437  & 0      \\
Mn     & 4  &  0   & 0  &  Mn-O   & 1345.15 &  0.324  & 0      \\
 O      & 0.862    &  -2.862  & 74.92 &  O-O  &  22764.3 &  0.149 & 27.88  \\
\end{tabular}\end{ruledtabular} \end{table*}

For the low-$T$  $A2_1am$ space group, the factor group analysis yields the total irreducible representation for Raman-active modes:
$\Gamma_R=18A_{1}(xx,yy,zz)+ 17A_2(xy) + 16B_1(xz) + 18B_2(yz)$. In this study, the intermediate phase is assumed to have the $Acaa$ symmetry. The total irreducible representation of Raman-active modes for the $Acaa$ space group  is given by $\Gamma_R=6A_{g}(xx,yy,zz)+ 11B_{1g}(xy) + 11B_{2g}(xz) + 11B_{3g}(yz)$. For the high-$T$ $I4/mmm$ space group, the Raman-active modes are
factored as $\Gamma_R=4A_{1g}(xx,yy,zz)+ B_{1g}(xx,yy) + 5E_{g}(xz,yz)$.

In order to assign the symmetries and eigenvectors to the observed phonon peaks, we computed the $\Gamma$-point phonon modes using shell-model lattice dynamical calculations implemented in the general utility lattice program (GULP) package. In the simple shell model,  an ion $Z$ consists of
a point core, carrying the total mass of the ion with charge $X$, and a massless shell with charge $Y$, representing the outer valence electrons.
The core and shell are coupled by a harmonic spring with a force constant $K$.
The parameter $K$ is related to the ionic polarizability $\alpha=Y^2/K$.
The interionic interactions between Ca, Mn, and O ions are described by a combination of long-range Coulomb potentials and short-range Born-Mayer-Buckingham potentials between ions $i$ and $j$:
 $$V_{\mathrm{BM}}(r)=A_{ij}\exp(-r/\rho_{ij})-C_{ij}/r^6,$$
where $A_{ij}$ and $\rho_{ij}$ denote the strength and the range of the repulsive interaction, respectively,
and $C_{ij}$ describes an attractive part with the interatomic distance $r$. With a well-documented set of data, the shell-model parameters are optimized to reach reasonable agreement with experimental Raman data~\cite{Lewis,Sherwood}. The resulting shell-model parameters are summarized in Table I.

\section{\label{sec:level3}Results and discussion}

\subsection{\label{sec:level3.1} First-order phonon excitations}
\begin{figure}
\label{figure1}
\centering
\includegraphics[width=9cm]{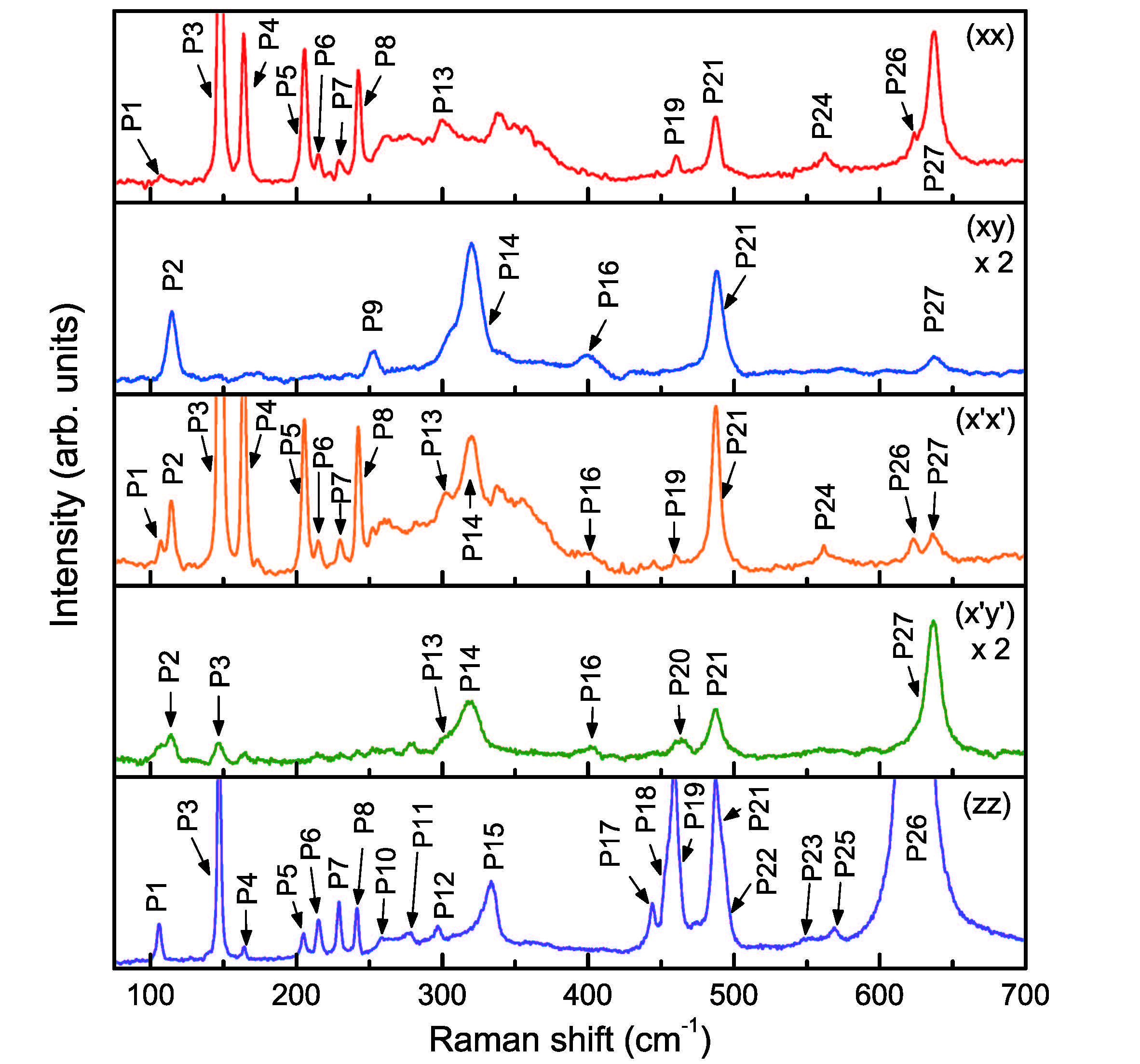}
\caption{Raman spectra of Ca$_3$Mn$_2$O$_7$ at an excitation wavelength of $\lambda=532$~nm in ($xx$), ($xy$), ($x'x'$), ($x'y'$), and ($zz$) polarizations measured at $T=8$~K. The most intense peaks are marked with indexes.}
\end{figure}
Figure~2 presents the polarized Raman spectra of Ca$_3$Mn$_2$O$_7$ measured at $T=8$~K  in five different [($xx$), ($xy$), ($x'x'$), ($x'y'$) and ($zz$)] polarizations. Here, $x$ and $y$ correspond to the directions along the Mn-O bonds,
and $z$ corresponds to the out-of-plane direction. $x'$ and $y'$ are rotated in-plane by 45$^{\circ}$ with respect to $x$ and $y$. These scattering geometries allow probing mainly  the $A_1$ and $A_2$ symmetry modes expected for the orthorhombic $A2_1am$ space group.

 Below 630~cm$^{-1}$ we were able to resolve ($15A_{1}$ + $4A_2$) one-phonon modes, which are smaller than the 35 Raman-active modes predicted for the $A2_{1}am$ structure in the selected polarizations,  $\Gamma=18A_1(xx, yy ,zz)+ 17A_2(xy)$. Assigning the observed phonon modes to the specific symmetry  is made on the basis of the polarization-dependent Raman spectra and the lattice-dynamical calculations
 with the implementation of the Bilbao Crystallographic Server~\cite{Kroumova}
 (see Sec. III and  Table II). The missing modes are ascribed to either a lack of phonon intensity or their overlap with other phonon excitations. In addition, we observe the weak symmetry-forbidden ($5B_{1}$ + $2B_2$) phonon modes possibly
due to either a leakage of a polarizer or a slight tilting of the aligned crystal toward the $c$ axis.  The spectra were fitted using a set of Lorentz functions. The extracted frequencies of the Raman-active optical phonon modes are listed in Table II. The calculated and experimental phonon energies agree well with each other. The representative
displacement patterns for the $A_{1}$ modes are sketched in Fig.~3.

\begin{figure}
\label{figure1}
\centering
\includegraphics[width=9cm]{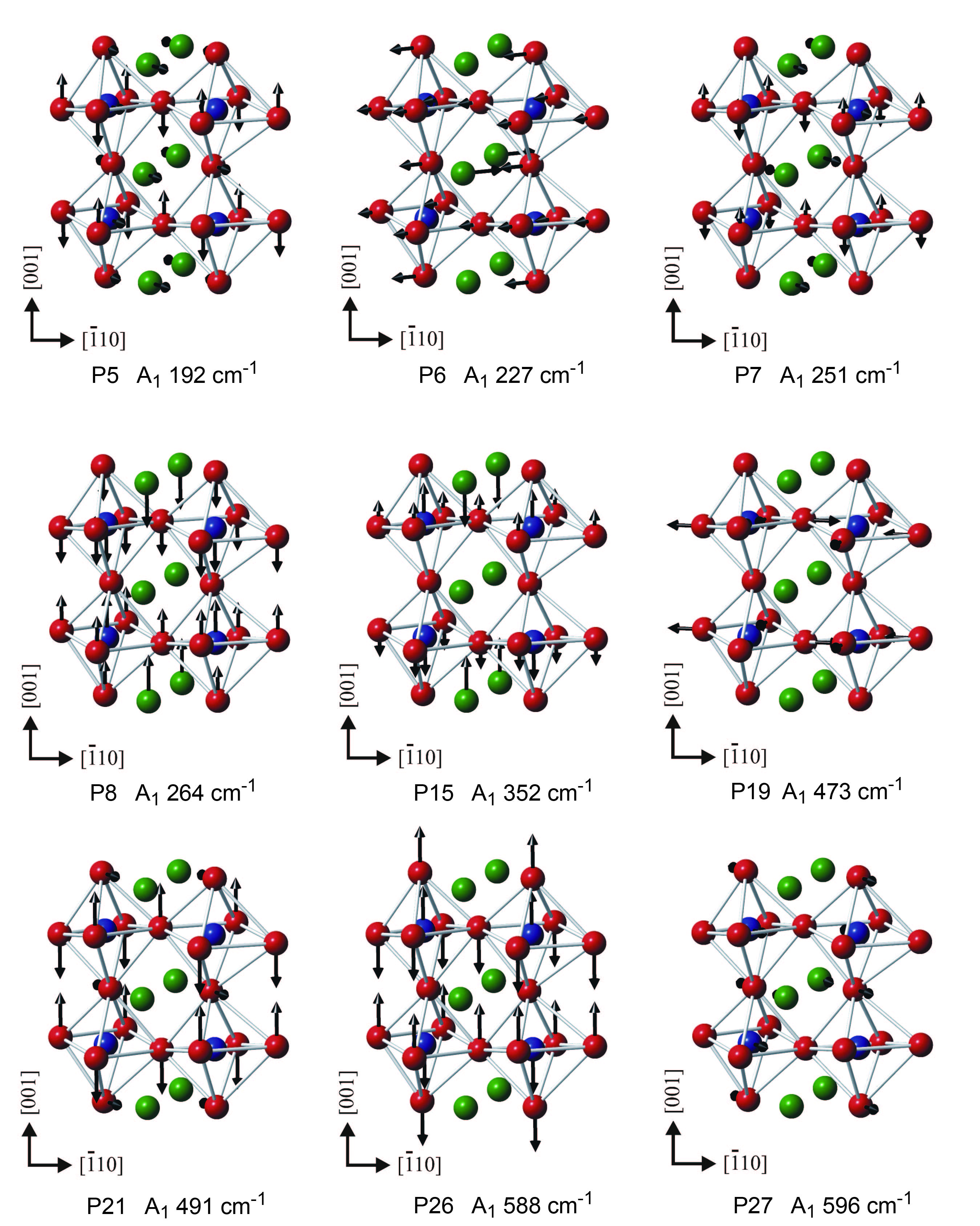}
\caption{Schematic representation of calculated eigenvectors for the $A2_1am$ symmetry. The relative amplitude of the vibrations is represented
by the arrow length. The blue, red, and green balls are Mn, O, and Ca atoms, respectively.  The numbers are the  calculated
frequencies using shell-model lattice dynamical calculations.}
\end{figure}

\begin{table*} \caption{\label{tab:table1} List of phonon symmetries and calculated frequencies in cm$^{-1}$ at the $\Gamma$ point for
Ca$_3$Mn$_2$O$_7$ with $A2_1am$ symmetry.}
\begin{ruledtabular} \begin{tabular}{ ccccc }
Mode & Exp. & Cal.  &  Selection rule &  Assignment  \\ \hline
P1  & 107 &	 113	&  ($xx$),($x'x'$),($zz$)	& A$_1$ (MnO$_6$ out-of-phase tilting; Ca(2) in-phase motions in the $ab$ plane)  \\
P2  & 114 &	133	&   ($xy$),($x'x'$),($x'y'$)	& A$_2$ (MnO$_6$ bending in the $ab$ plane)  \\
P3  & 150 &	136	&  ($xx$),($x'x'$),($x'y'$),($zz$)	& A$_1$ (MnO$_6$ in-phase rotation about the $c$ axis)   \\
P4	& 164 &	172	&  ($xx$),($x'x'$),($zz$)	& A$_1$ (MnO$_6$ in-phase bending; Ca(1)+Ca(2) out-of-phase motions along the $a$ axis)   \\
P5	& 205 &	192	&  ($xx$),($x'x'$),($zz$)	& A$_1$ (MnO$_6$ bending; Ca(1)+Ca(2) in-phase motions along the $b$ axis)   \\
P6	& 215 &	227	&  ($xx$),($x'x'$),($zz$)	& A$_1$ (MnO$_6$ stretching; Ca(1) motions in the $ab$ plane)   \\
P7	& 230 &	251	&  ($xx$),($x'x'$),($zz$)	& A$_1$ (MnO$_6$ out-of-phase rotation; Ca(1)+Ca(2) stretching in the $ab$ plane)   \\
P8	& 242 &	264	&  ($xx$),($x'x'$),($zz$)	& A$_1$ (MnO$_6$ stretching; Ca(2) out-of-phase motions along the $c$ axis)   \\
P9	& 253 &	274	&	 ($xy$)		& A$_2$ (MnO$_6$ stretching; Ca(1)+Ca(2) stretching along the $c$ axis);     \\
    &     &  276  &   ($xy$)              &     A$_2$ (MnO$_6$ bending; Ca(2) stretching along the $b$ axis)   \\
P10 &	260	& 282	 &		($zz$)	 &  A$_1$ (MnO$_6$ in-phase bending; Ca(1)+Ca(2) in-phase motions in the $ab$ plane)   \\
P11 &	278 & 300  &	   ($zz$)  & 	B$_2$ (MnO$_6$ in-phase bending; Ca(1)+Ca(2) in-phase motions in the $ab$ plane)   \\
P12	& 297	& 314	 &		($zz$)	&  A$_1$ (MnO$_6$ in-phase bending; Ca(2) stretching motions in the $ab$ plane)   \\
P13	& 303	& 328	 &  ($xx$),($x'x'$),($x'y'$)	&	A$_1$ (MnO$_6$ in-phase bending  in the $ab$ plane)   \\
P14	& 320	& 332 &     ($xy$),($x'x'$),($x'y'$)    &  A$_2$ (MnO$_6$ in-phase bending in the $ab$ plane)   \\
     &      & 344	&	 ($xy$),($x'x'$),($x'y'$) 	 & A$_2$ (MnO$_6$ in-phase bending in the $ab$ plane; Ca(1) stretching in the $ab$ plane)   \\
P15	& 338	& 352	&   ($zz$)	& A$_1$ (MnO$_6$ out-of-phase stretching; Ca(2) out-of-phase motions along the $c$ axis)   \\
P16	& 400	& 410	&	 ($xy$),($x'x'$),($x'y'$) 	& A$_2$ (MnO$_6$ bending; Ca(1) motions along the $b$ axis)   \\
P17 & 444	& 469 &	     ($zz$)				 &	B$_1$ (MnO$_6$ and Ca(2) stretching vibrations)   \\
P18	& 454	& 470	&	 ($zz$)	       	& B$_1$ (MnO$_6$ and Ca(2) stretching motions in the $ab$ plane)   \\
P19 & 459	& 473	& ($xx$),($x'x'$),($zz$)		& A$_1$ (MnO$_6$ stretching in the $ab$ plane)   \\
P20	& 465	&		&		($x'y'$)		    & two-phonon mode    \\
P21	& 488	& 491	& ($xx$),($xy$),($x'x'$),($x'y'$), ($zz$) & A$_1$ (MnO$_6$ bending )   \\
P22	& 492	& 502	&	($zz$)	& B$_2$ (MnO$_6$ bending; Ca(1)+Ca(2) out-of-phase stretching)  \\
P23	& 551	& 525	 &  ($zz$)	& B$_1$ (MnO$_6$ bending in the $ab$ plane)   \\
P24	& 562	& 556 &       ($xx$),($x'x'$)    &  B$_1$ (MnO$_6$ bending )   \\
     &       & 571 &	 ($xx$),($x'x'$)       & B$_1$ (MnO$_6$ bending; Ca(2) out-of-phase motion along the $a$ axis)   \\
P25	 & 568	 & 581  &	($zz$)				  &  B$_1$ (MnO$_6$ bending along the $b$ axis; Ca(2) out-of-phase motion along the $b$ axis)   \\
P26	 & 623	 & 588  &	 ($xx$),($x'x'$),($zz$) 	&   A$_1$ (MnO$_6$ out-of-phase bending in the $bc$ plane)   \\
P27	 & 637	 & 596 & 	 ($xx$),($xy$),($x'x'$),($x'y'$) 	  &  A$_1$ (MnO$_6$ bending in the $ab$ plane; Ca(1)+Ca(2) in-phase stretching)  \\
\end{tabular} \end{ruledtabular} \end{table*}

\begin{figure}
\label{figure1}
\centering
\includegraphics[width=8cm]{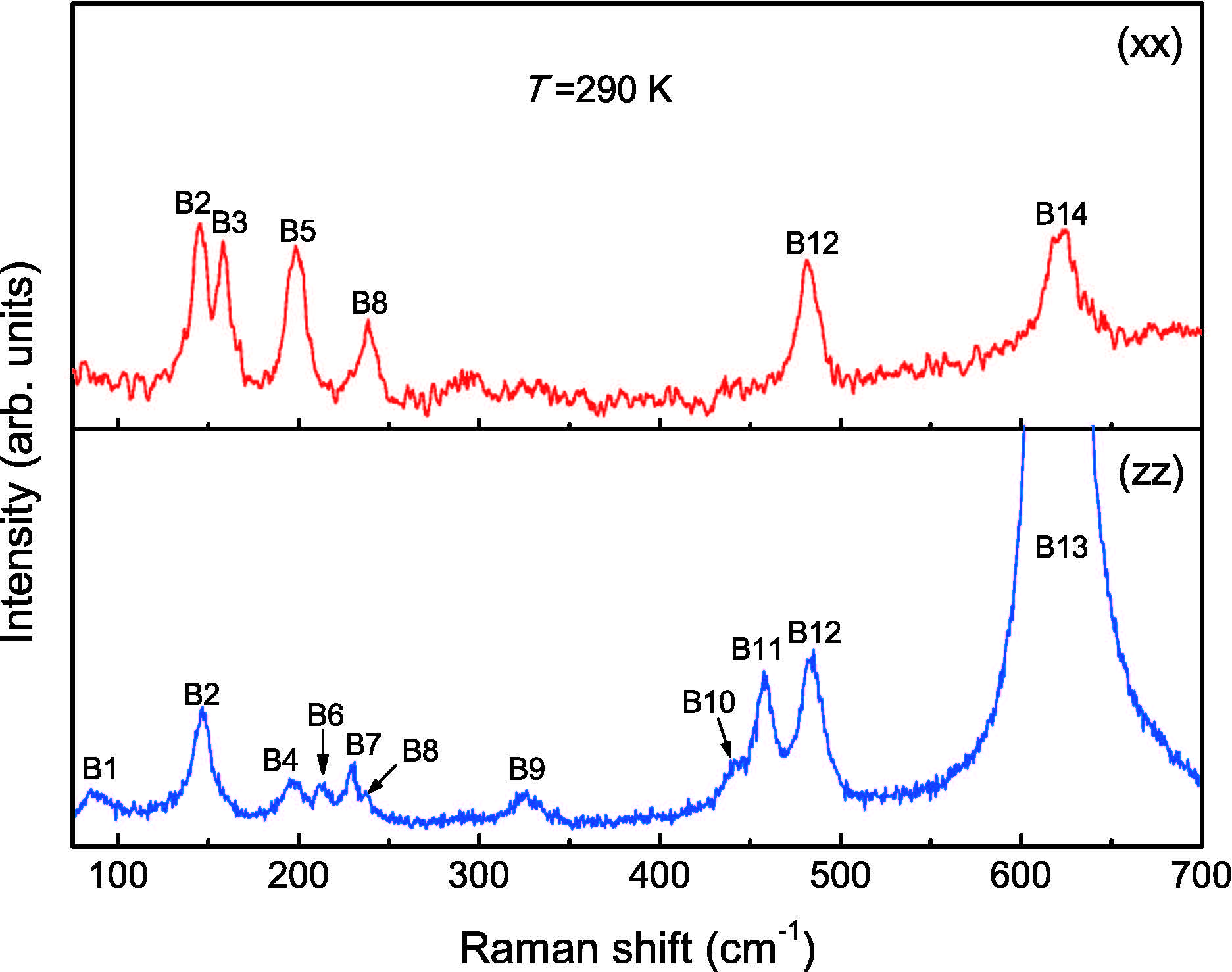}
\caption{Raman spectra of Ca$_3$Mn$_2$O$_7$ at an excitation wavelength of $\lambda=532$~nm  measured in ($xx$) and ($zz$) polarizations at $T=290$~K. The intense peaks are marked with index. }
\end{figure}

We next turn to the polarized Raman spectra of the intermediate phase recorded in ($xx$) and ($zz$) polarizations at $T=290$~K.
As shown in Fig.~4,  we observe 14 one-phonon modes. The observed phonon number is between the 18 Raman-active modes
$\Gamma_R=18A_{1}(xx,yy,zz)$ predicted for the $A2_1am$ symmetry and the 6 Raman-active modes $\Gamma_R=6A_{g}(xx,yy,zz)$ predicted for the $Acaa$ space group.
This result is consistent with the coexistence of the low-$T$ ($A2_1am$) and the intermediate-$T$ orthorhombic ($Acaa$) phases at room temperature~\cite{Senn15,Lobanov,Gao}.
Admittedly,  it is almost impossible to assign unambiguously the observed phonon modes according to their symmetries  as
the phonon energies anticipated for the $A2_1am$ and the $Acaa$ symmetries are nearly degenerate.

For ease of discussion, the phonon modes are tentatively described in terms of the $Acaa$ symmetry. As listed in Table III, we identify the ($5A_g$ + $3B_{1g}$ + $5B_{2g}$) modes. We sketch the displacement patterns of the representative normal modes in Fig.~5. In this assignment, the symmetry-forbidden ($3B_{1g}$ + $5B_{2g}$)  modes can arise from local lattice distortions or a small mismatch between the laboratory and the crystal coordinates. In reality, the forbidden modes may originate from the $A_1$ modes of the low-$T$ $A2_1am$  phase, which persist to the intermediate-$T$
phase.

\begin{figure}
\label{figure1}
\centering
\includegraphics[width=8.5cm]{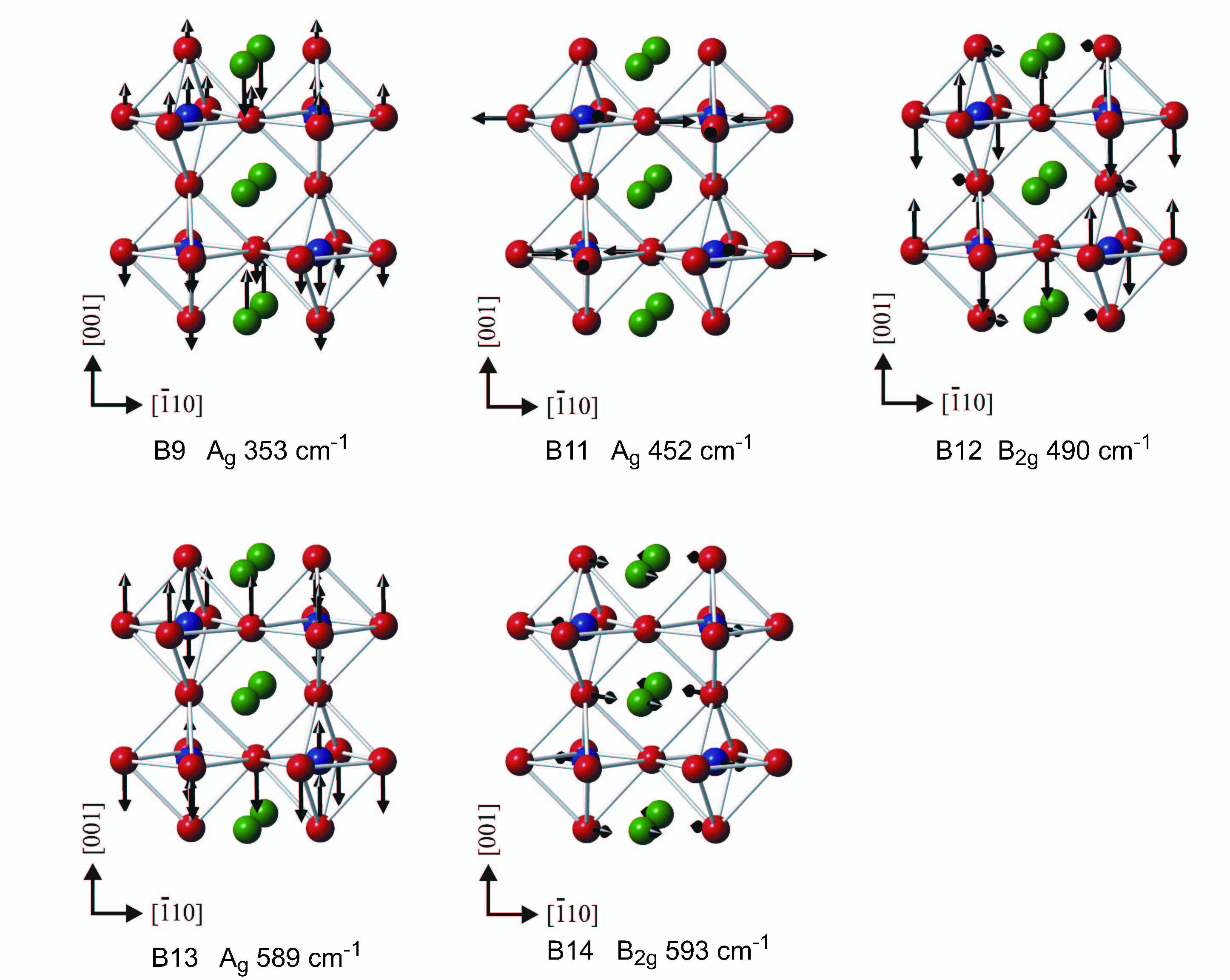}
\caption{Schematic representation of the calculated modes for the $Acaa$ symmetry. The numbers are the calculated frequencies. The related amplitude of the vibrations is represented by the arrow length. The blue, red, and green balls are Mn, O, and Ca atoms, respectively. }
\end{figure}

\begin{table*} \caption{\label{tab:table1} List of phonon symmetries and calculated frequencies in cm$^{-1}$ at the $\Gamma$ point for
Ca$_3$Mn$_2$O$_7$ with $Acaa$ symmetry.}
\begin{ruledtabular} \begin{tabular}{ ccccc }
Mode & Exp. & Cal.  &  Selection rule &  Assignment  \\ \hline
B1  & 85 &	 99	&  ($zz$)	& B$_{2g}$ (MnO$_6$ out-of-phase tilting; Ca(1)+Ca(2) stretching vibrations in the  $ab$ plane)  \\
B2  & 146 &	135	&  ($xx$),($zz$)				& A$_g$ (MnO$_6$ out-of-phase rotation about the $c$ axis)  \\
B3  & 158 &	151	&  ($xx$)	& B$_{1g}$ (MnO$_6$ out-of-phase rotation about the $c$ axis; Ca(2) stretching along the $b$ axis)  \\
B4	& 198 &	213	&  ($zz$)	& B$_{2g}$ (MnO$_6$ bending in the $ab$ plane; Ca(2) stretching  along the $a$ axis)   \\
B5	& 199 &	213	&  ($xx$)	& B$_{1g}$ (MnO$_6$ bending in the $ab$ plane; Ca(2) stretching  along the $b$ axis)  \\
B6	& 213 &	215	&  ($zz$)	& B$_{1g}$ (MnO$_6$ bending; Ca(1)+Ca(2) in-phase motion along the $a$ axis)   \\
B7	& 229 &	251	&  ($zz$)	& B$_{2g}$ (MnO$_6$ bending; Ca(1)+Ca(2) out-of-phase motion along the $b$ axis)   \\
B8	& 240 &	258	&  ($xx$),($zz$)	& A$_{g}$ (MnO$_6$ tilting; Ca(1)+Ca(2) out-of-phase motion along the $a$ axis)   \\
B9	& 325 &	353	&  ($zz$)	& A$_g$ (MnO$_6$ out-of-phase stretching; Ca(2) out-of-phase motion along the $c$ axis)   \\
B10 & 441	& 	 &		($zz$)	 &  mode of the $A2_1am$ space group \\
B11 & 458 & 452  &	   ($zz$)  & 	A$_g$ (MnO$_6$ out-of-phase stretching motions in the $ab$ plane)   \\
B12	& 483	& 490	 &  ($xx$),($zz$)	&  B$_{2g}$ (MnO$_6$ bending vibrations)   \\
B13	& 619	& 589	 &  ($zz$)	&	A$_g$ (MnO$_6$ out-of-phase bending in the $bc$ plane)   \\
B14	& 621	& 593 &     ($xx$)    &   B$_{2g}$ (MnO$_6$ bending in the  $ab$ plane; Ca(1)+Ca(2) in-phase stretching) \\
\end{tabular} \end{ruledtabular} \end{table*}

\subsection{\label{sec:level3.1} Higher-order Raman scatterings}

\begin{figure*}[tb]
\label{figure2}
\centering
\includegraphics[width=16cm]{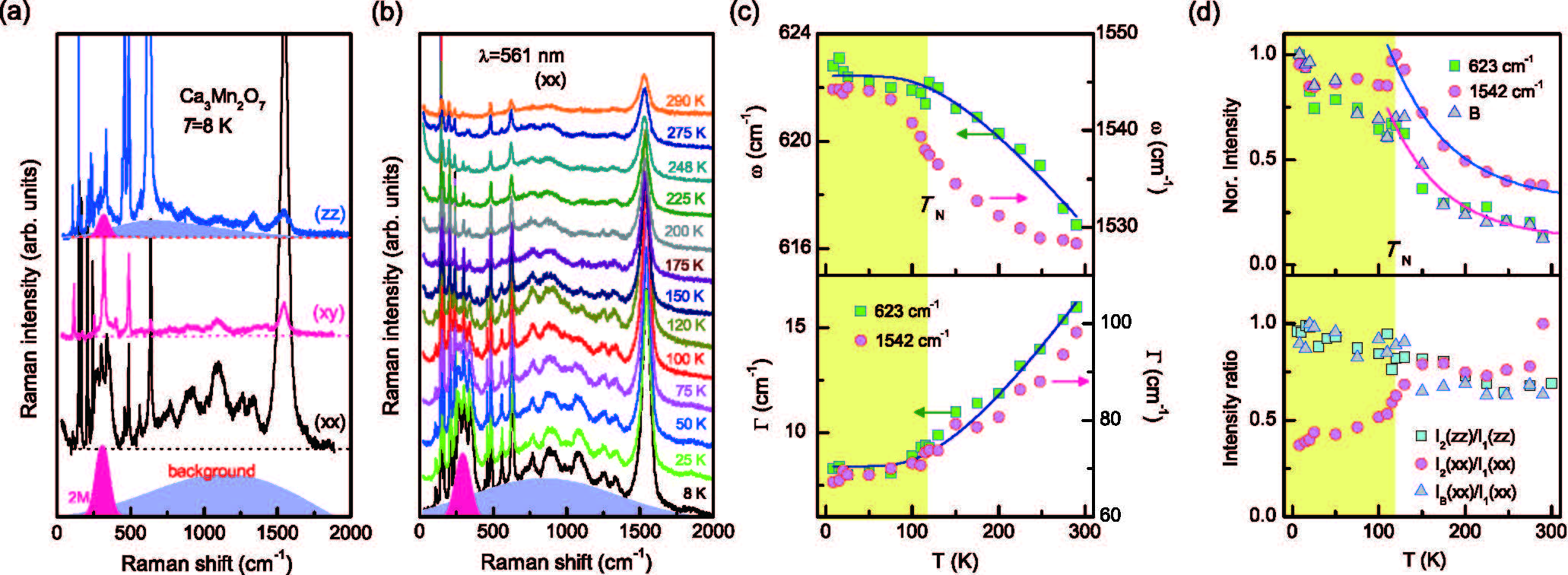}
\caption{(a) Raman spectra of Ca$_3$Mn$_2$O$_7$ measured at $T=8$~K in the ($xx$), ($xy$), and ($zz$) polarizations. The spectra were recorded
with an excitation wavelength of  $\lambda=532$~nm. The pink and purple shaded areas denote a two-magnon excitation and a background continuum, respectively. (b) Temperature dependence of the Raman spectra
in the ($xx$) polarization. The spectra are vertically shifted by a constant amount. (c) Temperature dependence of  the frequency and the linewidth of the one-phonon mode at 623~cm$^{-1}$ and hybrid excitation at 1542~cm$^{-1}$. The solid lines are fits to the anharmonic model. (d) Normalized scattering intensity of the phonons and background continuum  as well as their intensity ratios as a function of temperature and polarization. The solid lines are fits of the intensity using the
exponential form $I(T)\propto\exp(-k_BT/\Delta)$.}
\end{figure*}

We now focus on the higher-order Raman spectra. Figure~6(a) shows the $T=8$ K Raman spectra  of Ca$_3$Mn$_2$O$_7$
taken in a wide frequency range of $30 - 2000~\mbox{cm}^{-1}$ for ($xx$), ($xy$), and ($zz$) polarizations.
In addition to the one-phonon modes, we observe the two-magnon (2M) excitation (pink shading) and a dozen higher-order scatterings, which are
superimposed on top of the broad background (gray shading). It is remarkable that the higher-order mode at 1542 cm$^{-1}$
is much more intense than the one-phonon peaks only in the ($xx$) polarization and the background continuum is anisotropic in its intensity and center position between the ($xx$) and ($zz$) polarizations.

Figure 6(b) exhibits the $T$ dependence of the ($xx$) polarized Raman spectra.
With increasing temperature, all Raman excitations are drastically suppressed.
To quantify their temperature evolution, the Raman spectra are
fitted to a sum of Lorentzian profiles. Shown in Fig. 6(c) are the $T$ dependences of the frequencies $\omega$ and full widths at half maximum $\Gamma$ for the representative one-phonon mode at 623 cm$^{-1}$ and the anomalous higher-order peak at 1542 cm$^{-1}$. We find that both
$\omega(T)$ and $\Gamma(T)$ of the  623 cm$^{-1}$ mode, involving stretching vibrations of the MnO$_6$ octahedra, are well described by an anharmonic model [see the solid lines in Fig. 6(c)].  Unlike the one-phonon modes,  $\omega(T)$ of the higher-order 1542~cm$^{-1}$ mode  shows a steeper decrease than what is expected for the lattice anharmonicity, starting at $T_\mathrm{N}$. However,  $\Gamma(T)$ exhibits no apparent anomaly.

In Fig.~6(d) we compare the $T$ dependence of the normalized intensities of the one-phonon mode, the higher-order peak, and the background continuum. They commonly display an exponential-like decrease $I(T)\propto\exp(-k_BT/\Delta)$ as the temperature is raised above $T_\mathrm{N}$ [see the solid lines in Fig. 6(d)].  In insulating materials, the phonon intensity $I(T)$ is highly susceptible to changes in the dielectric function
with respect to the displacement of the normal mode. In contrast, the variation of penetration depth and scattering volume with temperature gives minor contributions. Thus, the exponential drop of $I(T)$ above $T_\mathrm{N}$ suggests that the dielectric function strongly varies as soon as the magnetic order disappears. Noticeably, the $X^{-}_3$ and $X^{+}_2$ modes start to soften at $T_\mathrm{N}$, and the extracted empirical energy $\Delta=95.8 \pm 1.3$~K  is comparable to the frequency of the soft T mode
(see Sec. IV. E and Fig.~9 below).

In the bottom panel of Fig.~6(d), we present the ratio $I_2/I_1$ of the higher- to first-order scattering intensity.
For the ($xx$) polarization,  $I_2/I_1$  undergoes a steplike increase with increasing temperature through $T_\mathrm{N}$, while for the ($zz$) polarization, it decreases monotonically. The opposite polarization dependence of $I_2/I_1$ indicates that the higher-order scattering arises from a
resonant optical excitation of 2.2 eV (see the following Sec. IV. C)
and that an electric state is anisotropic between the in- and out-of-plane directions.
A closer look at the higher-order modes reveals
that an integer multiple or combination of first-order phonon energies cannot reproduce higher-order peaks.
For instance,  the energy of the 1542 cm$^{-1}$ peak corresponds to 2.47 times the one-phonon frequency at 623 cm$^{-1}$.
This points towards a hybrid nature of the 1542 cm$^{-1}$ peak. In HIFs, higher-order coupling of multiple degrees of freedom can lead to the vibronic couplings of the multiphonon scatterings to magnon and/or soft phonons.  This together with
strong lattice instabilities may explain the broad (two)-phonon background ranging from 100 to $2000~ \mbox{cm}^{-1}$ in terms of a (two)-phonon density of states. The selective temperature, polarization, and incident laser dependences  exclude fluorescence as its origin.

\subsection{\label{sec:level3.1} Resonance Raman scattering}
\begin{figure*}
\label{figure1}
\centering
\includegraphics[width=14cm]{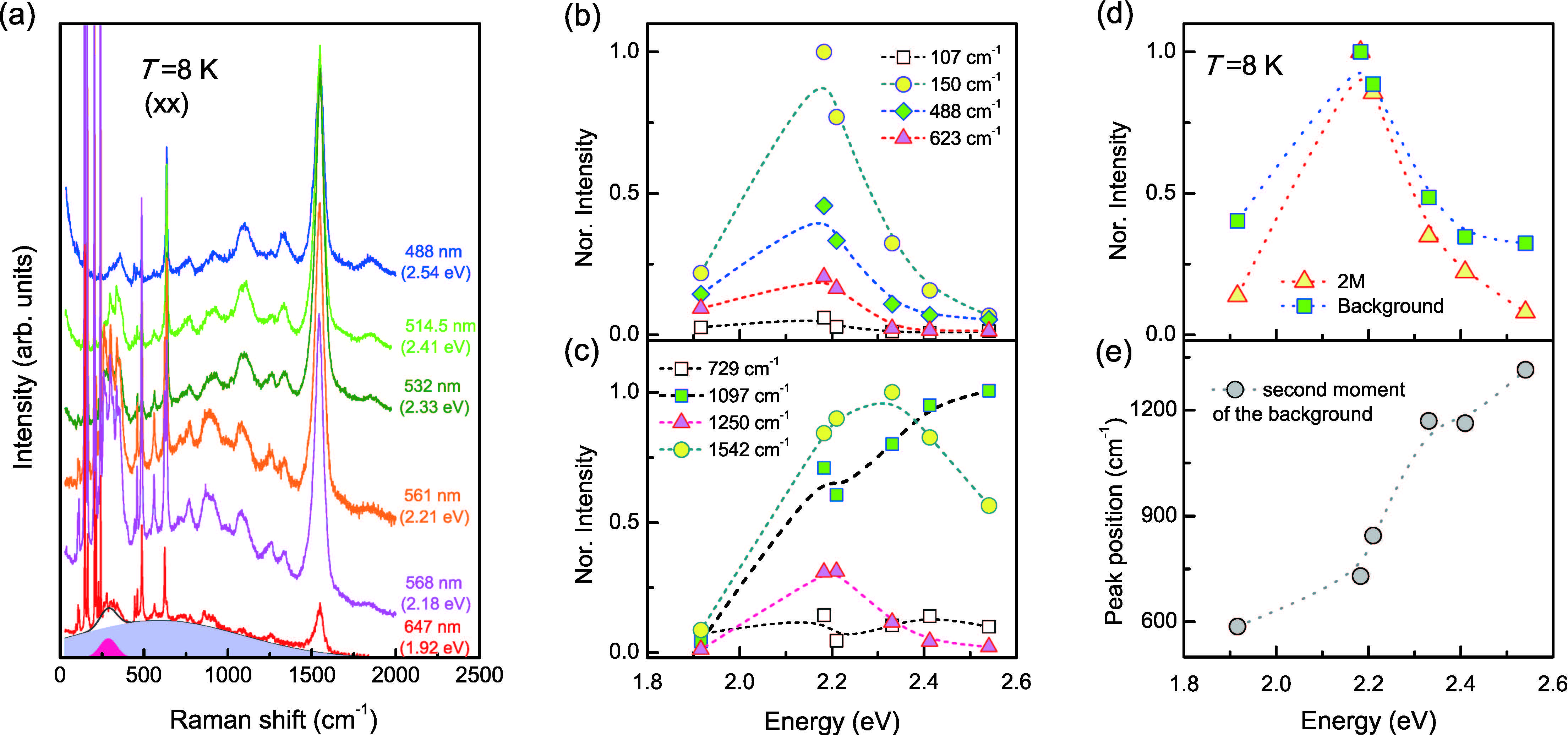}
\caption{(a) Resonant Raman spectra of Ca$_3$Mn$_2$O$_7$ measured at $T=8$~K in the ($xx$) polarization. The excitation energy is varied from 1.92 to 2.54 eV. The spectra are vertically shifted for clarity. (b) Normalized Raman scattering intensity of the first-order phonon modes as a function of the incident phonon energy. The dashed lines represent guides to the eye. (c)  Normalized Raman scattering intensity of the higher-order peaks as a function of the energy excitation. (d) Wavelength dependence of the normalized intensity of the magnetic excitation (yellow triangles) and the background continuum (green squares). (e) Second moment of the background continuum versus the laser excitation energy.}
\end{figure*}

We next discuss resonance behaviors of Ca$_3$Mn$_2$O$_7$. This is a particularly relevant issue because lattice instabilities can give rise to substantial modifications of electronic band structures. Figure 7(a) displays the resonant Raman spectra of Ca$_3$Mn$_2$O$_7$ measured with different wavelengths at $T=8$~K in the ($xx$) polarization. A large variation of the scattering intensity is observed as a function of the incident laser energies. For an analysis of the resonant behavior, all phonon and magnetic excitations were fitted using a sum of Lorentzian and Gaussian profiles. The resulting intensities versus incident laser energy are plotted in Figs.~7(b)-7(d).

As shown in Fig.~7(b), almost all the first-order phonon modes at 107, 150, 488 and 623 cm$^{-1}$ are resonant with the excitation energy at around 2.2 eV. This is contrasted by the resonant behavior of the higher-order modes; although the 1250 cm$^{-1}$ peaks show a resonant profile similar to that of the one-phonon peak, the 729, 1097 and 1542 cm$^{-1}$ peaks  exhibit two selective enhancements at about 2.2 and 2.54 eV [see Fig.~7(c)]. The observed resonant scatterings are ascribed to the spin-allowed electronic transition of Mn$^{4+}$ ion, $^4$A$_{2g}$($^4$F)-$^4$T$_{2g}$($^4$F), the $d-d$ transition $^4$A$_{2g}$($^4$F)-$^4$T$_{1g}$($^4$F) and the charge transfer transition O$^{2-}$($2p$)$\rightarrow$ Mn$^{4+}$($3d^3$)~\cite{Balykina,Brik}. Figure 7(d) displays the resonant profiles of the magnetic and background continua, which show essentially the same behavior as the one-phonon one. Noticeably, the second moment of the background continuum increases appreciably with increasing excitation energy, while undergoing a steep jump at about 2.2 eV [see Fig.~7(e)].  Overall, our resonant scattering study suggests that the higher-order scatterings are much more selectively enhanced at the 2.54 eV electronic transition than at the 2.2 eV one.

\subsection{\label{sec:level3.1}Two-magnon scattering}
\begin{figure*}
\label{figure1}
\centering
\includegraphics[width=14cm]{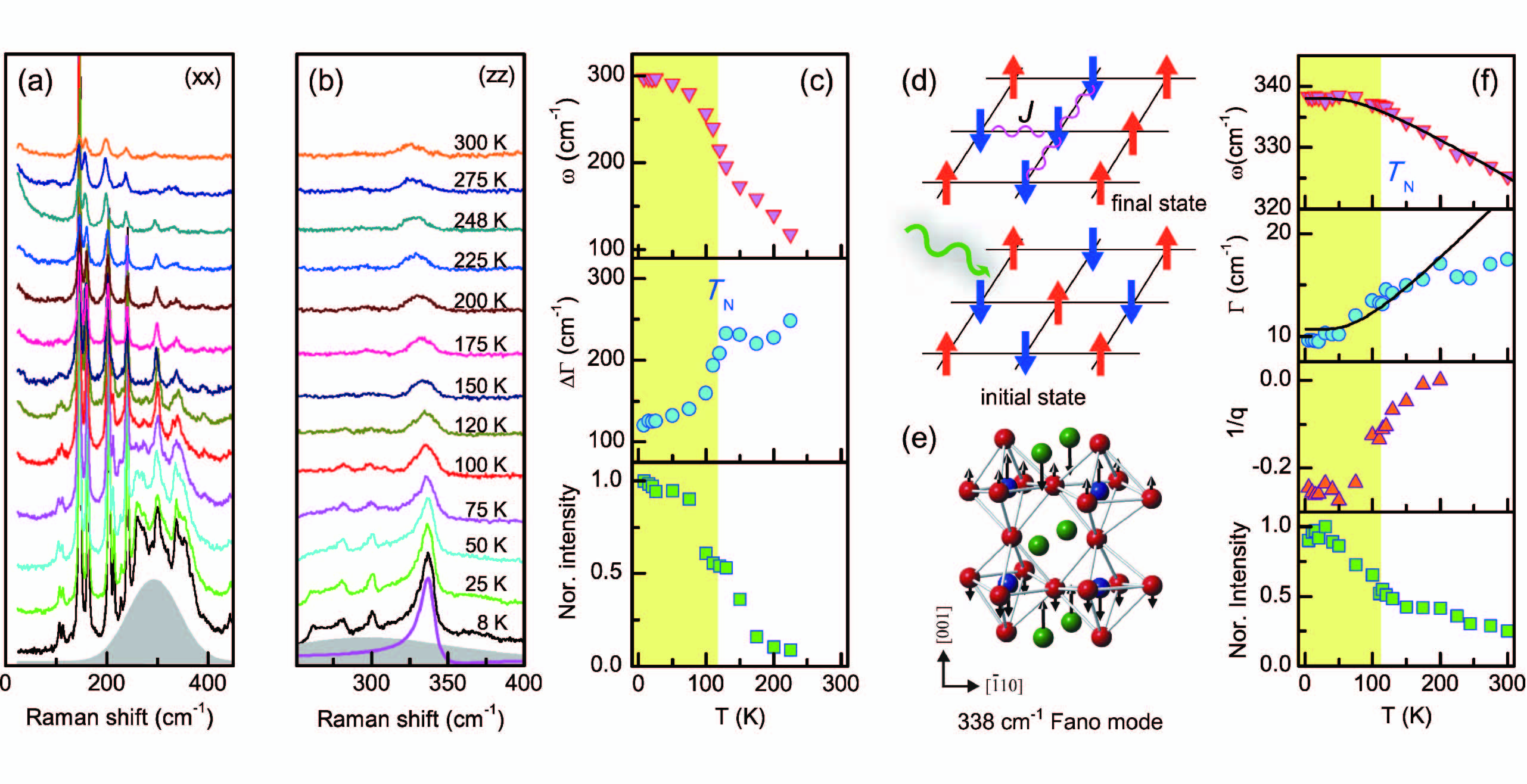}
\caption{(a),(b) Evolution of two-magnon excitation (gray shading) in the ($xx$) and ($zz$) polarizations. The purple solid line is a Fano resonance.
(c) Temperature dependence of the frequency, linewidth, and intensity of two-magnon scattering. The yellow shading marks $T_\mathrm{N}$.
(d) Cartoons of the two-magnon Raman scattering process. (e) Schematic representation of the eigenvector of the 338~cm$^{-1}$ $A_1$ symmetry mode. (f) Temperature dependence of the
frequency, linewidth, Fano asymmetry, and normalized intensity. The solid
lines are fits using an anharmonic model. }
\end{figure*}

We inspect the 2M scattering arising from double spin-flip processes of a ground
state ($S=3/2, S_z=\pm 3/2$) to a higher state ($S=3/2, S_z=\pm 1/2$)  [see the sketch in Fig.~8(d)]~\cite{Cottam}. As shown in Fig.~8(a), the 2M excitation is observed as a Gaussian-like maximum at about 301~cm$^{-1}$
in the ($xx$) polarization. The magnetic excitation is strongly suppressed with increasing temperature through $T_\mathrm{N}$.
Figure 8(c) exhibits the $T$ dependence of the 2M frequency, FWHM, and intensity. On approaching $T_\mathrm{N}$,
the magnon (2M peak) energy is renormalized by 30 \% and the magnon lifetime ($\Gamma$) becomes two times shorter.
Compared to the two-dimensional $S=2$ system LaSrMnO$_4$~\cite{Choi}, Ca$_3$Mn$_2$O$_7$ experiences a stronger thermal damping and renormalization of the magnon
despite its apparent smaller spin number. In addition, the 2M scattering does not evolve to a pronounced quasielastic response above $T_\mathrm{N}$ unlike
conventional 2M excitations, lacking well-defined paramagnons at high temperatures. These anomalies point to the presence of another relaxation channel,
possibly due to inherent coupling of the magnetic subsystem to the soft tilt mode. The peak position $\omega_\mathrm{2M}$ of the 2M scattering allows the estimation of the exchange coupling constant
$J_\mathrm{nn}$ of the nearest-neighbor Mn spins by the relation $\omega_\mathrm{2M}=J_\mathrm{nn}(2zS-1)$, where $z=5$ is the number of nearest-neighbor spins and $S=3/2$ is the spin number. The value of $\omega_\mathrm{2M}=301$~cm$^{-1}$ yields $J_\mathrm{nn}=30.9$~K. This is close to the theoretical value of $J=39$~K~\cite{Matar}.

Remarkably, the 2M scattering is also observed with  weaker intensity in the ($zz$) polarization
as zoomed in Fig.~8(b). Further evidence for a magnetic continuum is provided by a Fano resonance of the 338-cm$^{-1}$ mode.
The appearance of the 2M excitations in the out-of-plane polarization is totally unexpected for the tetragonal K$_2$NiF$_4$-type antiferromagnets~\cite{Choi}
and is linked to the CaMnO$_3$ bilayer structure, leading to significant interlayer interactions. Indeed, Ca$_3$Mn$_2$O$_7$ has the dominant G-type antiferromagnetic ordering of the Mn$^{4+}$ magnetic moments pointing to the $z$ axis~\cite{Lobanov}. The 338 cm$^{-1}$  $A_1$ symmetry mode involves out-of-phase vibrations of the MnO$_6$ octahedra and the Ca(2) atoms along the $z$ axis, as sketched in Fig.~8(e). This mode is fitted using the Fano profile, $I(\omega)=I_0[1+(\omega-\omega_0)/q\Gamma)^2]/[1+((\omega-\omega_0)/\Gamma)^2]$, where
$I_0$, $\omega_0$, $\Gamma$, and $1/|q|$ are the intensity, the bare phonon frequency, the linewidth, and the asymmetric parameter, respectively. In Fig.~8(f), all Fano parameters
are plotted as a function of temperature. The $T$ dependence of $\omega_0$  is well
described by the conventional anharmonic model. However,  $\Gamma(T)$ deviates for temperatures above 200~K.
On heating above 50 K, $1/|q|$ decreases rapidly toward zero, while the intensity falls off substantially up to $T_\mathrm{N}$ and then decreases gradually.
The Fano-parameter behaviors lend further support to a quick damping of the magnons above  $T_\mathrm{N}$.

\subsection{\label{sec:level3.1} Soft modes}

\begin{figure*}
\label{figure1}
\centering
\includegraphics[width=16cm]{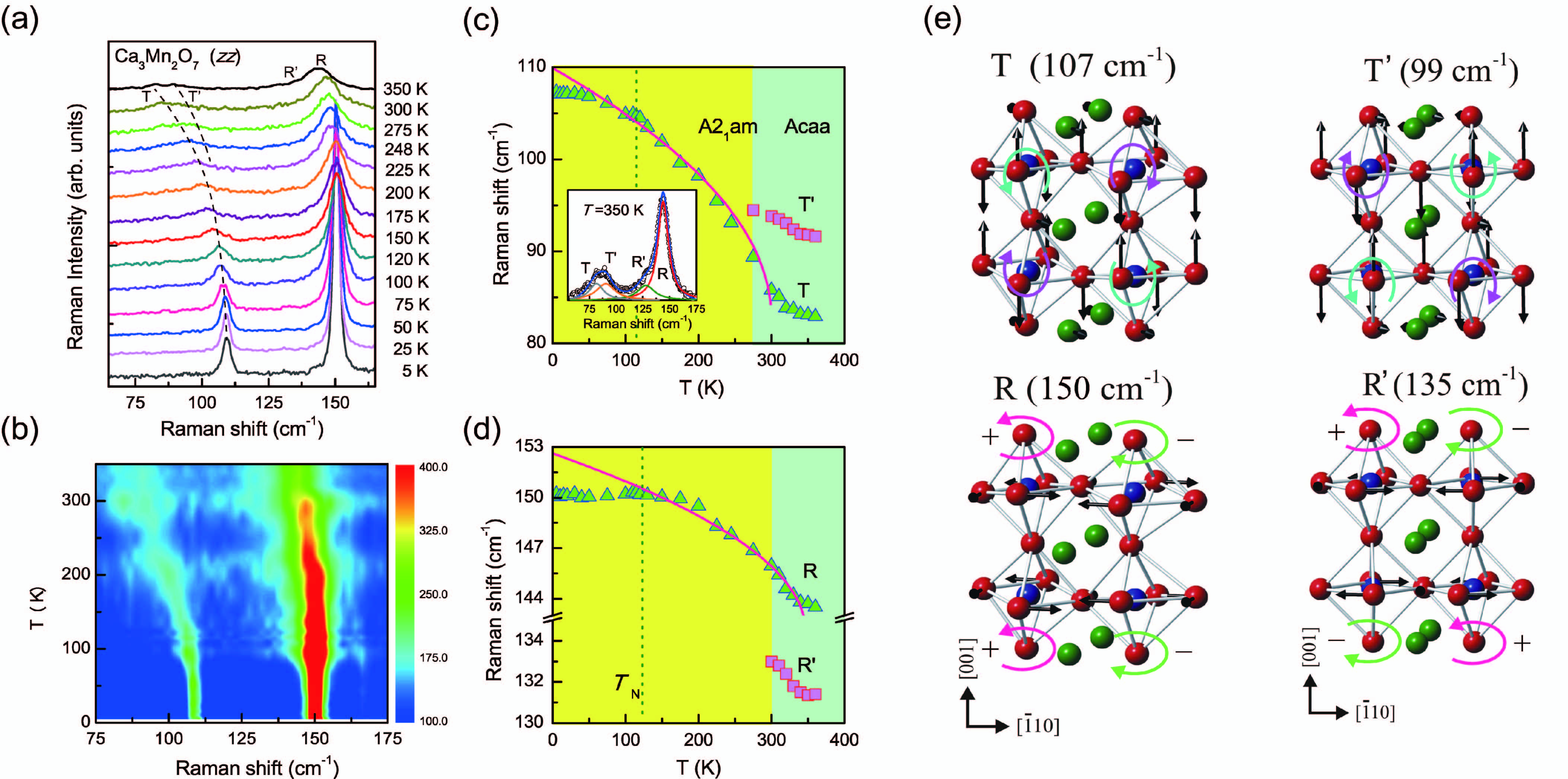}
\caption{(a) Temperature dependence of the low-frequency Raman spectra ranging from 60 to 160~cm$^{-1}$ in the ($zz$) polarization. T and T' (R and R') denote the soft tilt (rotational) modes. (b) Color contour plot of the soft-mode intensity
in the temperature-Raman-shift plane. (c) and (d) Temperature dependence of the frequencies of the soft modes. The solid lines
are fittings using the mean-field model described in the text. The inset shows a representative fit of the
low-energy spectrum to four Lorentzian profiles at $T =350$~K, representing the T, T', R and R' soft modes. (e) Schematic representation of eigenvectors of the soft modes
in the $Acaa$ and $A2_1am$ phases.}
\end{figure*}

Figure~9(a) zooms in on the two low-frequency phonons at 107 (T) and 150~cm$^{-1}$ (R) measured on warming.
The data were taken in the ($zz$) polarization in order to separate the T and R modes from other phonon peaks.
Based on our lattice dynamical calculation, the 107~cm$^{-1}$ (150~cm$^{-1}$) phonon is assigned to the octahedron tilting (rotation) mode whose normal mode displacements are sketched in Fig.~9(e).

With increasing temperature toward the $Acaa$ phase,
the T (R) mode displays a huge (small) softening  by 26 (6)~cm$^{-1}$. For temperatures above 270~K, the T and R modes are split into additional
T' and R' modes, forming the coexistence regime of the two different orthorhombic phases at least up to 360~K [see the inset of Fig.~9(c)].
This together with the drastic softening of the T mode confirms
that the first-order phase transition from the $A2_1am$  ground state to the intermediate  $Acaa$
phase is driven by the tilting of oxygen octahedra. Essentially the same conclusion was drawn
in a recent neutron and x-ray diffraction study of Ca$_3$Mn$_{1.9}$Ti$_{0.1}$O$_7$ that unravels
the softness of the antiphase octahedral tilt and robustness of the octahedral rotation~\cite{Ye}.
The selective susceptibility of the antiphase tilt mode to the structural transformation
is discussed in terms of the decisive role of the partially occupied $d$ orbital of the Mn$^{4+}$ ions
in stabilizing the MnO$_6$ distortion. In Fig.~9(b) we show the color contour plot
of the soft-mode intensity in the temperature-Raman-shift plane. Upon approaching
the structural phase transition, the soft-mode intensities are strongly suppressed.

The frequencies $\omega_\mathrm{S}(T)$ of the T and R modes are shown in Figs.~9(c) and 9(d) in the temperature range of $T=5 -360$~K.
Notably,  the softening modes do not shift to zero frequency,  unlike the zone-center polar soft mode in a proper ferroelectric transition.
$\omega_\mathrm{S}(T)$ is described by the mean-field formula
$\omega_S(T)=A(T_\mathrm{S}-T)^{1/2}+\omega_\mathrm{O}$. Here, the second constant term accounts for the first-order nature of the structural phase transition.
We find a good agreement between the experimental data
and the mean-field theory above $T_\mathrm{N}$, confirming the soft-mode driven first-order phase transition.
The apparent deviation in the antiferromagnetic ordered state is
attributed to strong spin-lattice coupling $\lambda$, leading to a renormalization
of the phonon energy by a scalar spin-correlation function $\Delta\omega=\lambda \langle \mathbf{S}_i\cdot \mathbf{S}_j\rangle$.

Last, we pay attention to an alteration of the rotation and tilting displacement patterns through the
phase transition from the $Acaa$ to the $A2_1am$ symmetries.
Overall, the energies of the tilting and rotational modes between the two phases differ by $12- 17~\mbox{cm}^{-1}$.
This is due to a slightly different pattern of the MnO$_6$ tilting and the  rotation distortions between them.
The T and T' modes involve the respective inward  and outward
tilting of the MnO$_6$ octahedra within the perovskite bilayer, while the R and R' modes correspond to the $X^{+}_2$ in-phase and  $X^{-}_1$ out-of-phase rotations of the MnO$_6$ octahedra within the perovskite bilayer, respectively [see Fig.~9(e)]. We stress that the intermediate phase is characterized by the coexistence
of the soft modes with distinct displacements, which is responsible for the observed non-switchable polarization and uniaxial negative thermal expansion~\cite{Gao,Senn15,Senn16}.
In addition, the competing structural instabilities at the intermediate phase lead to a rapid destabilization of
the magnetic excitation, in disfavor of strong magnetoelectric coupling. To achieve simultaneous ferroelectricity and magnetoelectricity, thus,
the patterns of octahedral tilt and rotation distortions should be finely controlled through a structural transformation.

\section{\label{sec:level5}Conclusion}
In conclusion, our Raman scattering study of Ca$_3$Mn$_2$O$_7$
has unveiled a set of competing soft modes, which dictate a first-order
phase transition from the paraelectric to the ferroelectric orthorhombic phase.
The soft rotation and tilt modes are a tuning parameter of all
physical properties, leading to pronounced higher-order scatterings,
anomalous magnetic excitation, and structural instability.
In spite of the intertwined coupling among lattice, magnetic, and structural
degrees of freedom, the soft modes with different symmetries
prevent strong magnetoelectric coupling and ferroelectric switching.
Our work demonstrates that competing octahedral tilting and rotational distortions created through a phase transition
should be avoided to achieve room-temperature multiferroics.

\section*{\label{sec:level6}Acknowledgments}

This work was supported by Korea NRF Grants (No. 2009-0093817 and 2017-012642), DFG Le967/16-1, DFG-RTG 1953/1, Metrology for Complex Nanosystems, and
the NTH-School Contacts in Nanosystems: Interactions, Control and
Quantum Dynamics, and the Gordon and
Betty Moore Foundation¡¯s EPiQS Initiative through Grant No. GBMF4413 to the Rutgers Center for
Emergent Materials. The crystal growth at Rutgers was supported by the NSF MRI Grant No. MRI-1532006.


\end{document}